\documentclass[useAMS,usenatbib,a4paper,fleqn,letter]{mn2e}

\usepackage{amsfonts,amssymb,amsmath}
\usepackage{aas_macros}
\usepackage{times,varioref,multirow,textcomp,comment}
\usepackage{xspace,float}
\usepackage[usenames,dvipsnames,svgnames,hyperref]{xcolor}
\usepackage[pdftex]{graphicx}
\usepackage[percent]{overpic}
\usepackage{}
\usepackage{comment}
\usepackage[
    pdftex,
    a4paper=false,
    plainpages=true,
    pdfpagelabels,
    breaklinks=true,
    bookmarks=true,
    bookmarksopen=false,
    bookmarksopenlevel=2
    bookmarksnumbered=true,
    bookmarkstype=toc,
    colorlinks=true,
    citecolor=RoyalBlue,
    linkcolor=ForestGreen,
    menucolor=Teal,
    urlcolor=DarkOrange,
]{hyperref}

\providecommand{\adsurl}[1]{\href{#1}{ADS}}

\def \Msun{\ {\rm M_\odot}}

\def \Mpch{\ h^{-1}{\rm Mpc}}

\def \LCDM{$\Lambda$CDM}

\def \vmax{V_{\rm max}}

\newcommand{\Eqref}[1]{Eq.~(\ref{#1})}
\newcommand{\Figref}[1]{Fig.~\ref{#1}}
\newcommand{\Secref}[1]{\S\ref{#1}}

\defcitealias{nfw}{NFW}

\defcitealias{elahi2011}{{\sc stf}}

\defcitealias{ahf}{{\sc ahf}}

\def \Gadget2{{\sc gadget-2}}

\hypersetup{
  pdfauthor = {Pascal Elahi},
  pdfkeywords = {Galaxy Groups, Velocity Dispersion},
  pdftitle = {Group Velocity Dispersions}
}

\def\surfs{{\sc surfs}}

\begin{document}
\title[Galaxy Group Velocity Dispersions]{Using Velocity Dispersion to Estimate Halo Mass: Is the Local Group in Tension with $\Lambda$CDM?}
\author[P.J.~Elahi, et al. ]{
\parbox{\textwidth}{
Pascal J. Elahi$^{1,2}$\thanks{E-mail: pascal.elahi@icrar.org},
Chris Power$^{1}$,
Claudia del P. Lagos$^{1}$,
Rhys Poulton$^{1}$,
Aaron S.G. Robotham$^{1}$,
}\vspace{0.4cm}\\ 
\parbox{\textwidth}{
$^{1}$International Centre for Radio Astronomy Research, University of Western Australia, 35 Stirling Highway, Crawley, WA 6009, Australia\\
$^{2}$ARC Centre of Excellence for All Sky Astrophysics in 3 Dimensions (ASTRO 3D)\\
}
}
\maketitle

\pdfbookmark[1]{Abstract}{sec:abstract}
\begin{abstract}
    Satellite galaxies are commonly used as tracers to measure the line-of-sight velocity dispersion ($\sigma_{\rm LOS}$) of the dark matter halo associated with their central galaxy, and thereby to estimate the halo's mass. Recent observational dispersion estimates of the Local Group, including the Milky Way and M31, suggest $\sigma\sim$50 km/s, which is surprisingly low when compared to the theoretical expectation of $\sigma\sim$100s km/s for systems of their mass. Does this pose a problem for $\Lambda$CDM? We explore this tension using the {\small{SURFS}} suite of $N$-body simulations, containing over 10000 (sub)haloes with well tracked orbits. We test how well a central galaxy's host halo velocity dispersion can be recovered by sampling $\sigma_{\rm LOS}$ of subhaloes and surrounding haloes. Our results demonstrate that $\sigma_{\rm LOS}$ is biased mass proxy. We define an optimal window in $v_{\rm LOS}$ and projected distance ($D_p$) -- $0.5\lesssim D_p/R_{\rm vir}\lesssim1.0$ and $v_{\rm LOS} \lesssim0.5V_{\rm esc}$, where $R_{\rm vir}$ is the virial radius and $V_{\rm esc}$ is the escape velocity -- such that the scatter in LOS to halo dispersion is minimised - $\sigma_{\rm LOS}=(0.5\pm0.1)\sigma_{v,{\rm H}}$. We argue that this window should be used to measure line-of-sight dispersions as a proxy for mass, as it minimises scatter in the $\sigma_{\rm LOS}-M_{\rm vir}$ relation. This bias also naturally explains the results from \cite{mcconnachie2012a}, who used similar cuts when estimating $\sigma_{\rm LOS,LG}$, producing a bias of $\sigma_{\rm LG}=(0.44\pm0.14)\sigma_{v,{\rm H}}$. We conclude that the Local Group's velocity dispersion does not pose a problem for $\Lambda$CDM and has a mass of $\log M_{\rm LG, vir}/M_\odot=12.0^{+0.8}_{-2.0}$. 
\end{abstract}
\maketitle

\section{Introduction}\label{sec:intro}
Currently favoured theories of galaxy formation predict that galaxies are embedded within massive haloes of non-baryonic Cold Dark Matter (CDM) \cite[e.g.][]{white1991a,baugh2006,benson2010b}. These haloes play a fundamental role in regulating galaxy properties, as is evident in scaling relations such as, for example, those between stellar mass and halo mass \cite[e.g.][]{guo2010a,behroozi2010a,reddick2013a}. While factors such as a halo's assembly history and its larger-scale environment will influence galaxy properties, there are sound physical arguments as to why a halo's mass should be particularly important. The dark matter mass governs the depth of the gravitational potential well within which galaxies evolve, and it impacts directly the timescales on which galaxies grow, via gas accretion and mergers, and the efficiency with which feedback from stars and black holes influences gas dynamics within the galaxy \cite[e.g.][]{white1991a}. 

\par
Testing these ideas observationally requires accurate estimates of halo mass that can be determined on a system-by-system basis, which favours the use of satellite galaxies as dynamical tracers. On galaxy cluster mass scales, numerous observational estimators have been published, but it has been shown that velocity dispersion measurements of cluster galaxies in the highest mass systems allow the mass to be recovered with the smallest bias, less than $\sim$0.2~dex \cite[e.g.][]{old2014a,old2015a}. Using velocity dispersions on galaxy group mass scales is more challenging because individual systems tend to contain fewer dynamical tracers, i.e. satellites, and estimates are likely to be more uncertain. Nevertheless, this approach has been used successfully by \citet[e.g.][]{carlberg1997b,schneider2006,yang2006,robotham2011a}. 

\par
Biases arise invariably because assumptions about a halo's dynamical state and geometry are unavoidable. A typical halo is assumed to be (almost) spherical and virialised, with a satellite population drawn from a dynamically relaxed distribution, but haloes as they might exist are likely to be triaxial spheriods whose axis ratios axis depend on halo mass \cite[e.g.][]{elahi2018a} with a virialisation state that depends on mass and environment. Major mergers significantly affect a halo's dynamical state, but its average growth history is dominated by smooth mass accretion and minor mergers \cite[see][]{rodriguezpuebla2016a,elahi2018a}. Consequently, using a population of satellites as tracers of the dynamical mass of an object will give reasonable, though possibly biased, mass estimate. 

\par 
Interestingly, the Local Group (LG) has a velocity dispersion of $\sim 50$ km/s \cite{mcconnachie2012a}, much lower than we might expect given the expected mass of the system (see \Figref{fig:groupdispersion}). Our local group is a the system in which we might expect the biases arising from incomplete tracer populations, missing satellites below some magnitude limit, to be minimal. Does this low dispersion pose a problem for $\Lambda$CDM, imply that the LG is unusual, or might it reflect uncertainties in the mass of, for example, the Milky Way (MW), which could be low? The mass of MW is a contentious issue, with some arguing for high masses of $\sim1.5\times10^{12}\Msun$ \cite[e.g.][]{besla2007a,phelps2013a,reid2016a,zaritsky2017a,mcmillan2017a}, while other data argues for lower masses of $\sim0.8\times10^{12}\Msun$ \cite[e.g.][]{watkins2010,kafle2014a,huang2016a,eadie2016a}. Some of the disagreements likely relates to differences in tracers used, assumptions made and methods used \cite[see][for a summary of several methods used for the MW]{courteau2014a}, while some, particularly for MW, has to do with differences in probed radii and hence mass enclosed \cite[][]{eadie2017a}. 

\par 
Here, we seek to answer what is the bias for using the velocity dispersions of satellites around a central and whether the Local Group is unusual. Specifically, what biases are introduced given a satellite selection? Where does the scatter originate from? Can we improve the selection criteria in LOS velocities and distances to minimise the bias? Our paper is organised as follows: we discuss our data in \Secref{sec:data}, present our results in \Secref{sec:results} and end with a discussion of how to incorporate the bias and scatter to produce robust results. 

\section{Numerical Data}\label{sec:data}
\begin{table*}
\setlength\tabcolsep{2pt}
\centering\footnotesize
\caption{Halo Properties: We list the number of primary haloes, the virial mass of the host haloes $M_{\rm vir}$, the number of candidate haloes within two virial radii, the number of subhaloes, the mass \& lifetime of the surrounding haloes. For each quantity we characterise the distribution with (min,median$^{+1\sigma}_{-1\sigma}$,max)}
\begin{tabular}{l|cc|lcccc}
\hline
\hline
    Simulation & \multicolumn{2}{c}{Haloes} & \hspace{2pt} & \multicolumn{4}{c}{Surrounding (sub)haloes} \\
    & $N_{\rm main}$  & $M_{\rm vir}$ ($\log\Msun$) &
    & $N(r<2R_{\rm vir})$ & $N_{\rm sub}$ 
    & $M_{\rm vir}$ ($\log\Msun$) & Lifetime (Gyrs) \\
\hline
    L40N512     & 527   
    & $(11.6,12.1^{+0.5}_{-0.2},14.3)$ &
    & $(1,10^{+15}_{-6},1302)$        & $(0,7^{+15}_{-5},809)$ 
    & $(7.8,10.1^{+0.6}_{-0.4},14.3)$  & $(0.47,11.8^{+0.8}_{-2.7},13.3)$
    \\
    L210N1024   & 3959
    & $(12.1,13.3^{+0.4}_{-0.2},14.9)$ &
    & $(1,13^{+20}_{-7},834)$        & $(0,8^{+13}_{-5},280)$ 
    & $(9.0,11.3^{+0.6}_{-0.4},14.8)$ & $(0.46,10.4^{+1.7}_{-5.7},13.2)$ \\
    L210N1536   & 13551 
    & $(11.13,12.8^{+0.5}_{-0.2},14.9)$ &
    & $(1,12^{+24}_{-7},2741)$        & $(0,7^{+15}_{-4},763)$
    & $(8.5,10.7^{+0.6}_{-0.3},14.9)$ & $(0.46,10.4^{+1.5}_{-6.6},13.0)$ \\
\end{tabular}
\label{tab:groups}
\end{table*}

We use the \surfs\ simulations \cite[][]{elahi2018a}, a suite of N-body simulations of volumes ranging from $40\Mpch$ to $900\Mpch$, each containing billions of particles, run assuming a \LCDM\ Planck cosmology with $\Omega_{\rm M}=0.3121$, $\Omega_{\rm b}=0.0459$, $\Omega_{\mathrm \Lambda}=0.6879$, a normalisation $\sigma_8=0.815$, a primordial spectral index of $n_{\rm s}=0.9653$, and a Hubble parameter of $h_0=0.6751$ \cite[cf. Table of][]{planckcosmoparams2015}. We use a memory lean version of {\small GADGET2} \cite[][]{gadget2}, storing 200 snapshots evenly spaced in logarithm of the expansion factor between $z=24$ to $z=0$ to accurately capture the evolution of dark matter haloes; this temporal spacing ensures that we can follow the freefall time of overdensities of $200\rho_{\rm crit}$, i.e., haloes. Halo catalogues are constructed with the {\small VELOCIraptor} phase-space halo finder \cite[][]{elahi2011,elahi2013a}.

\par 
We focus on the subset of \surfs\ simulations with box sizes $40\Mpch$ and $210\Mpch$ (cf. Table~\ref{tab:groups}) and between $\sim$0.1 and 3 billion particles. This provides us with a sufficient statistical sample of well resolved central haloes - essentially friends-of-friends groups - with virial masses of $\sim10^{12}\Msun$, and allows us to identify the host haloes of galaxies with stellar masses of $\sim10^{8}\Msun$\footnote{This estimate is based on extrapolating the stellar mass to halo mass relation \cite[e.g.][]{moster2010a,behroozi2010a,vanuitert2016a}}. Here we limit our analysis to well-resolved central haloes composed of $\gtrsim10^4$ particles, with subhaloes and moderately well resolved neighbouring central haloes composed of at least $50$ particles, approximately twice the particle limit at which haloes are identified\footnote{Haloes composed of fewer particles are more susceptible to artificially shortened lives because they (1) drop below the particle limit at  which haloes are identified; and (2) have artificially softened density profiles  because of gravitational softening and so are more prone to tidal disruption \cite[e.g.][]{vandenbosch2017a}.}; a summary of these ``central'' haloes and the tracer population is given in Table~\ref{tab:groups}. With these data, we can estimate reliably central halo velocity dispersion using both the full particle distribution and surrounding sub- and central haloes as discrete tracers of the velocity field. This catalogue spans haloes with virial mass from $10^{11.5}-10^{15}\Msun$, where we define virial mass as $M_{\rm vir}=4\pi R_{\rm vir}^3 \Delta \rho_{\rm crit}/3$, with $\Delta=200$, $\rho_{\rm crit}$ is the critical density of the universe. A typical group in our sample consists of $\sim10$ subhaloes (satellites) and a few nearby but distinct central haloes within $\sim1.5R_{\rm vir}$.

\par
The focus of this work is on orbits, specifically those in dark matter only simulations. These orbits are reconstructed using {\sc OrbWeaver}, a tool that comes the {\sc VELOCIraptor} package. This code uses the evolution of haloes from merger trees produced by {\sc TreeFrog}, the {\sc VELOCIraptor} merger tree builder (see \cite{elahi2018a} for details) and identifies orbits around candidate hosts by tracking objects from $\sim2R_{\rm vir}$ and identifying changes in the sign of the radial velocity as peri/apo centres (for more details please see \Secref{sec:orbweaver}). Approximately $5\%$ of the orbiting (sub)haloes in our sample are short-lived with lifetimes of $\lesssim1$~Gyrs, most ($\sim 85\%$) of which are poorly resolved with $\sim50$ particles. These short-lived haloes are typically newly formed rather than systems with incorrectly tracked orbits and artificially shortened life-times. Based on the reconstructed lifetimes, this orbit catalogue is $\sim97\%$ complete, i.e., only $5\%$ of subhaloes are short-lived and of those, only $5\%$ are well resolved and artificially short-lived.

\par 
This orbit catalogue is extracted from DM only simulations. The addition of baryons, star formation and associated feedback processes (e.g. supernovae, Active Galactic Nuclei) does alter the dark matter distribution. At low dark matter halo masses, the (sub)halo occupancy is not unity, i.e., there are dark subhaloes \cite[e.g.][]{sawala2015a,garrison-kimmel2017a,sawala2017a}. This typically occurs below halo masses of $\lesssim10^{9.5}\Msun$ or maximum circular velocities of $\lesssim25$~km/s, the exact scale dependent on the feedback model used. In general, we are looking at scales well above these rough thresholds and so our results are not strongly influenced by the reduced occupancy. Another baryonic affect to consider is subhalo survival. 

\par 
Hydrodynamical simulations have fewer surviving subhaloes than dark matter only counterparts, a consequence of the stronger tidal field near the central galaxy \cite[e.g.][]{garrison-kimmel2017a,sawala2017a}. This tidal field can reduce the total number of subhaloes within $100$~kpc of a $\sim10^{12}\Msun$ halo (near the virial radius) by a factor of 2. The disrupted satellites are more likely to be of the dark variety as subhaloes with stars are more concentrated and less prone to complete disruption. Again, we typically focus on larger subhaloes due to the resolution limits of our simulations. 

\par
Orbits themselves can be affected by baryons. \cite{barber2014a} found luminous satellites tended to occupy more radial orbits than the total subhalo population of a $\sim10^{12}\Msun$ halo, albeit using a semi-analytic model and orbits from a dark matter only simulation. Given halos are dominated by dark matter save in the central tens of kpc, we might expect pericentres, and hence ellipticities to change due to the presence of baryons. However, the ellipticity of orbits is not critical to this work and will not affect greatly our results. 

\section{Satellites \& halo mass estimates}\label{sec:results}
\begin{figure}
    \centering
    \includegraphics[width=0.49\textwidth]{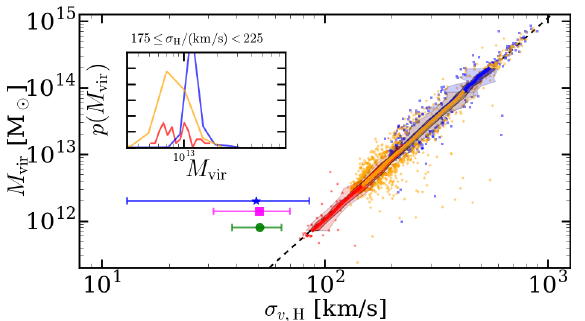}
    \caption{{\bf Halo Dispersions}. The mass of a halo given its dispersion from L40N512 (red), L210N1024 (blue) and L210N1536 (orange) simulations. We plot the median and $16\%,84\%$ quantiles as thick lines with contours for each simulation, outliers of the contour region and the tails of the mass distribution of well resolved haloes. We also show observational estimates of MW (green circle), M31 (magenta square), and the Local Group (blue star). Inset shows the distribution of mass within some dispersion range.}
    \label{fig:groupdispersion}
\end{figure}

We start by comparing the distribution of halo velocity dispersions, $\sigma_{v,{\rm H}}$, to halo masses in \Figref{fig:groupdispersion}; here velocity dispersion is measured using all particles within $R_{\rm vir}$. Our expectation is that
\begin{align}
  \sigma_{v,{\rm H}}\propto(\Delta/2)^{1/6}(H_0GM_{\rm vir})^{1/3},\label{eqn:sigmamass}
\end{align}
where $\Delta=200$ is the virial overdensity, $H_0$ is the Hubble constant, and $G$ is the gravitational constant. This is shown by the dashed black line in \Figref{fig:groupdispersion} and it provides a reasonable description of the simulation data; the simulated haloes shows little scatter with respect to this expectation, roughly $5\%$ independent of mass. The proportionality constant in \Eqref{eqn:sigmamass} is $0.68\approx2/3$.

\par 
However, the picture is not so simple when considering observational data, which must rely  on sparse sampling of the velocity distribution by using satellite galaxies residing in subhaloes  as tracers. This is challenging even with high quality spectroscopic data because the numbers of tracers are few.  Additionally, observationally assigning group/cluster membership is not a trivial, unlike in simulations where we make use of the full phase-space distribution to  separate the virialised phase-space envelop of a halo (central galaxy) and the subhaloes (satellite  galaxies) that reside in this region from other surrounding ``field'' haloes (galaxies).
\begin{figure}
    \centering
    \includegraphics[width=0.49\textwidth]{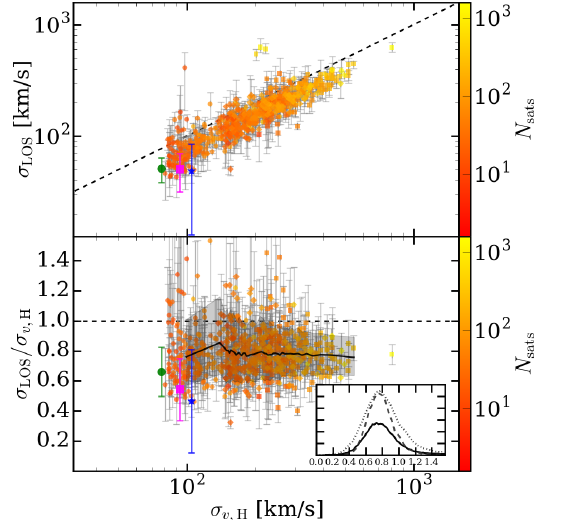}
    \caption{{\bf Observed Dispersions vs Halo Dispersions}. LOS velocity dispersion measured using tracers around a central halo along multiple lines-of-sight as a function of halo dispersion (top panel) and the ratio between these quantities (lower panel) for a random sample of 200 haloes. For each halo, we plot the median LOS dispersion along with error bars encapsulating the minimum and maximum values and colour code points according to number of satellites. In the lower panel we show the median by a solid black line and $16\%,84\%$ quantiles by shaded region for the entire halo sample, calculated in 10 bins with each bin containing $10\%$ of the population. The inset shows the distribution of ratios in three halo dispersion bins, solid line for the median$\pm1\sigma$, dashed (dotted) for high (low) dispersion systems, where we have folded in the scatter arising from varying lines-of-sight. We also show the one-to-one line as a dashed black line to guide the eye. We also place the MW, M31, and LG systems at halo dispersions based on the mean relationship between $M_{\rm vir}$ \& $\sigma_{v,{\rm H}}$ seen in \Figref{fig:groupdispersion}.}
    \label{fig:haloobsdisp}
\end{figure}

\par
A comparison of the true halo velocity dispersion and the observed velocity dispersion inferred from satellites, that is surrounding (sub)haloes, is presented in \Figref{fig:haloobsdisp}. Here we show the line-of-sight (LOS) velocity dispersion inferred from satellites, $\sigma_{\rm LOS}$, in our simulations relative to the true underlying halo dispersion $\sigma_{v,{\rm H}}$. We calculate the satellite dispersion using LOS motions of all haloes within a projected radius of $2R_{\rm vir}$ relative to the host halo in question, whether or not they are true subhaloes within the phase-space envelop or field haloes. We take several LOS and calculate the mean, maximum and minimum $\sigma_{\rm LOS}$. For clarity, we plot a subset of haloes, although the median, $1\sigma$ contours and histograms are derived from the entire population. We include also the scatter resulting from different LOS into the histograms and the $1\sigma$ contour. 

\par 
Figure~\ref{fig:haloobsdisp} highlights that $\sigma_{\rm LOS}$ typically underestimates $\sigma_{v,{\rm H}}$ across a wide range of halo masses by $\approx0.8\pm0.17$; even in rich clusters with lots of satellites, $\sigma_{\rm LOS}$ is biased. If we fit a Gaussian to the distribution of $\sigma_{\rm LOS}$, we find that even for large groups with $\sigma_{v,{\rm H}}\gtrsim250$km/s, the dispersion in this ratio is $0.15$. Limiting our LOS estimates to only those haloes within $1.25R_{\rm vir}$, we recover similar results. At face value, satellites appear to be biased tracers of the velocity field; although $\sigma_{\rm LOS}$ generally underestimates the true dispersion, it also overestimates the dispersions $15-20\%$ of cases. significant outliers are typically systems undergoing major mergers. Of particular concern is that the observed dispersion and mass estimates of MW, M31, and LG appear significantly colder than expected, a point which we will return to later. 

\par
First let us consider the average underestimate seen in \Figref{fig:haloobsdisp}, which could be a result of sparse sampling of the velocity field. We test how well the LOS velocity dispersions recover the true 3-dimensional halo dispersion using idealised N-body realisations of dark matter haloes following an NFW profile produced by {\sc galactICS} \cite[][]{widrow2008,galactics}. We do this by randomly sampling 2 million dark matter particle realisations 1000 times, each sample containing only 20 particles from which the LOS dispersion is calculated. We find that, in general, the LOS dispersion has no significant bias, with $\sigma_{\rm LOS}/\sigma_{H}=0.91^{+0.43}_{-0.28}$ for an NFW halo. This ratio is offset from that seen in \Figref{fig:haloobsdisp} and also has larger scatter, suggesting that our selected tracers, which contain both orbiting subhaloes, infallling haloes and neighbouring but unassociated haloes, do not perfectly trace the phase-space distribution of the central halo. 

\par
\begin{figure}
    \centering
    \includegraphics[width=0.49\textwidth]{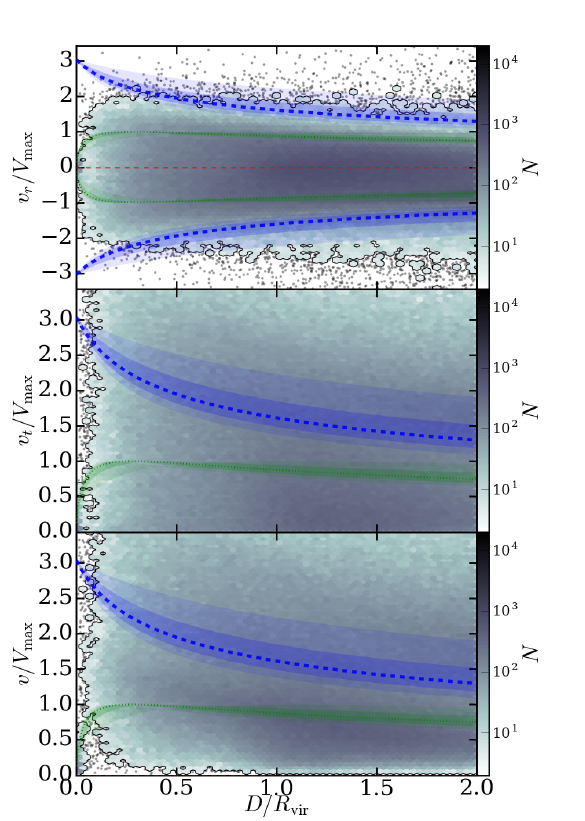}
    \caption{{\bf Motions of tracers}. We show stacked radial, tangential and total speed distribution as a function of distances for all haloes within 2.0 virial radii composed of $\gtrsim50$ particles for which we have well defined evolutionary tracks. For each halo we normalise the distance to group centre, $D$, by the virial radius $R_{\rm vir}$ and the velocities by the maximum circular velocity, $\vmax$. Outliers from the distribution are shown as small gray points. Dashed blue line shows the median escape velocity with the shaded region showing the variation in this limit from different halo concentrations, with the dark (light) showing the $1\sigma$ ($2\sigma$) contour. We also show the circular velocity by the dotted green line, with the associated shaded region giving the $1\sigma$ and $2\sigma$ contours.}
    \label{fig:satelliteorbits}
\end{figure}

 
We examine the motions of these candidate tracers in \Figref{fig:satelliteorbits}, where we show the distribution of radial, tangential and total velocities as a function of radial distance for all our haloes. We stack haloes by normalising distances and velocities by the virial radius $R_{\rm vir}$ and the maximum circular velocity $V_{\rm max}$ respectively. One of the notable features of this plots is the non-negligible fraction of candidate tracers that have velocities greater than the escape velocity, some of which reside well within the virial radius. The fraction of escaping subhaloes within one virial radius is $0.25\pm0.10$ (where the variance here is the halo-to-halo scatter). These subhaloes are not on bound orbits -- they may become backsplash subhaloes or leave entirely. Given the existence of a population of subhaloes on unbound orbits, the naive expectation would be for the LOS velocity based satellite dispersion to overestimate the halo dispersion, rather than underestimate it, if no cuts are applied to this data. 
\begin{figure}
    \centering
    \includegraphics[width=0.49\textwidth]{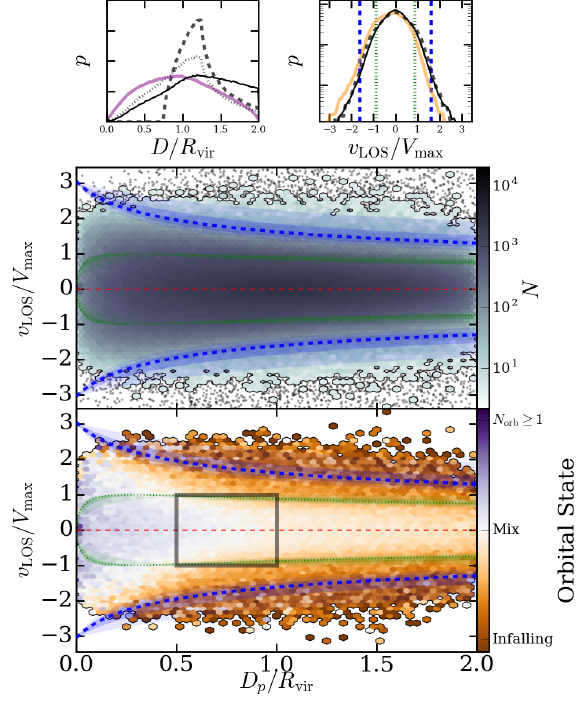}
    \caption{
    {\bf Line-of-sight orbit phase-space}. The {\em lower two panels} shows the LOS phase-space distribution of candidate satellite (sub)haloes, that is objects within 2$R_{\rm vir}$ of the halo. In the {\em middle panel} we show the number distribution of objects. The {\em bottom panel} we show the median orbital state at a given $(D_p,v_{\rm LOS})$ in hexagonal bins containing at least 5 (sub)haloes via the colour of the bin. We also show the optimal $\sigma_{\rm LOS}$ window (see text), outlined by a thick solid line in this panel. The {\em two upper panels} are histograms of the true radial distance distribution (upper left) and the LOS velocity distribution (upper right). {\em Upper left histograms are}: full 3D radial distribution (solid black line); projected radial distance distribution (thick purple line); resulting 3D radial distribution when a projected radial cut of $0.75\leq D_p/R_{\rm vir}\leq 1.25$ and $D_p/R_{\rm vir}\leq 1.25$ is applied (thick dashed and thick dotted line respectively). {\em Upper right histograms are:} the $v_{\rm LOS}$ distribution with $0.75\leq D_p/R_{\rm vir}\leq 1.25$ (solid line); the true radial velocity distribution within this projected radial cut (thick orange line); and $v_{\rm LOS}$ distribution of objects within $D<R_{\rm vir}$ in this projected radial cut window (thick dashed line), that is true subhaloes/satellites. We also show in this panel vertical dashed lines corresponding to the average circular velocity (green dotted) and escape velocity (blue dashed) to guide the eye.
    }
    \label{fig:satelliteorbitslos}
\end{figure}

\par 
However, this full three-dimensional information is not observationally accessible. Instead, observations must rely on LOS phase-space of haloes as presented in \Figref{fig:satelliteorbitslos}; that is, the projected radius $D_p$ and the LOS velocity $v_{\rm LOS}$ \cite[see for instance][for discussions and interpretations of LOS phase-space]{gill2005,oman2013a,oman2016b,jaffe2015a,yoon2017a}. Here the middle panel shows the number density distribution similar to \Figref{fig:satelliteorbits} and the bottom panel shows the mean orbital state. To calculate the mean orbital state, we use our orbit catalogue and identify objects on first infall (no change in sign) and those that have just completed first infall having passed pericentre (half an orbit). These are placed in the same category, the ``infalling class''. This class also includes interlopers, haloes that are never part of the host group halo with radially outgoing velocities. The other class of objects are those that have completed at least one full orbit. We calculate the mean class of haloes at a given $D_p$ \& $v_{\rm LOS}$ to determine the orbital state. 

\par 
The number density distribution (middle panel of \Figref{fig:satelliteorbitslos}) shows candidate tracers cover projected distances out to the virial radius. The LOS motion is centred on zero and extents out past the circular velocity threshold and even the escape velocity threshold. However, the distribution is peaked at low velocities with $\lesssim5\%$ of tracers on escape velocities with $D_p/R_{\rm vir}\leq1$, unlike the true value of $\sim25\%$. The projected radial distribution is also more centrally concentrated that the true underlying distribution (see probability distribution in upper-left panel, comparing the thick solid purple line to the solid black line). Overall, the $v_{\rm LOS}$ distribution lies primarily within the circular velocity envelop of a halo over a wide range of projected radii. Applying radial projection cuts can result in 3D radial distributions that differ from the true underlying distribution. An example of this is shown in the upper-right panel of \Figref{fig:satelliteorbitslos}, where we have selected haloes close to the virial radius in projected space (see thick dashed line in upper-left panel). Despite the distorted radial distribution, the velocity distribution within this projected radial cut is not appreciably different from that of haloes that are true radial distances of $\sim R_{\rm vir}$ (solid black line compared to solid orange line in upper-right panel), nor from orbiting haloes that are within the virial radius (solid black line compared to dashed black line in upper-right panel). 

\par 
The origin of the bias and the scatter seen in \Figref{fig:haloobsdisp} is seen in the lower panel of \Figref{fig:satelliteorbitslos}, where we plot the orbital state at a given LOS phase-space position. At moderate projected radial distances and $v_{\rm LOS}$, the halo population is dominated by objects on first infall and, as such, {\em will not trace the halo's phase-space distribution}. Objects that have completed at least one orbit and are part of the host halo's phase-space distribution and sample a wide range of velocities are concentrated to within $D_p\lesssim0.2R_{\rm vir}$. Uniformly sampling this LOS phase-space means including both real tracers of the halo's dispersion, i.e., subhaloes on bound orbits that have virialised, and interlopers, i.e., newly infalling and unassociated haloes. The result is the mild bias but large scatter seen in \Figref{fig:haloobsdisp}. 

\par 
Ideal tracers reside in the crowded central region; however, sampling this region has several issues. One is simply observational: placing slits to measure spectra in crowded regions and interpreting results is not trivial. The other issue is that objects within small projected radii actually span a large range in radial distances and thus do not sample the same velocity distribution. An example of such a cut is shown in the upper-left panel of \Figref{fig:satelliteorbitslos}, where we show the true radial distribution of all haloes identified with $D_p\lesssim R_{\rm vir}$ by a dashed black line. This selection has a significant fraction of haloes actually located at much larger radii. The observationally tractable outskirts contains a mix of infalling and orbiting haloes. Fortunately, in these outer regions, objects with high LOS motions are dominated by infalling haloes and those not associated with the central halo and only objects with $v_{\rm LOS}\lesssim V_{\rm circ}$ are a mix of orbital states. We argue that in order to minimise the dispersion in the true radial distances sampled and to minimise the contributions of interlopers (objects that have not even complete first pericentric passage) both a projected distance and LOS velocity cut needs to be applied. 

\par 
We search for this optimal window in a grid of projected radial windows centred on some $D_p$ with a width $\Delta D_p$ and maximum LOS velocity threshold $v_{\rm LOS,max}$ and identify the window with the best fitness. This fitness is defined as 
\begin{align}
    \mathcal{F}=&\left(1-\left|\frac{\mu_{D}-D_p}{D_p}\right|\right)\left(1-\left|\frac{\sigma_{D}-\Delta D_p}{\Delta D_p}\right|\right)\times\notag\\
    &\left(1-\left|\frac{v_{\rm esc}(D_p-\Delta D_p/2)-v_{\rm esc}(D_p+\Delta D_p/2)}{v_{\rm esc}(D_p)}\right|\right)\times\notag\\
    &\left(\frac{N_{\rm orb}}{N_{\rm int}}\right)
    \left(1-\frac{N_{\rm int}}{N_{\rm win}}\right)\times\notag\\
    &\left(\frac{N_{\rm win}}{N_{\rm all}}\right)\left(\frac{N_{\rm H}({N_{\rm sat}>=3})}{N_{\rm H}}\right)\times\notag\\
    &(1-|\delta{\sigma_{\rm LOS}}|).
    \label{eqn:fitness}
\end{align}
Here the first two terms are associated with the 3 dimensional radial distribution resulting from the projected distance cut. Ideally, this distribution should be similar to the projected one, therefore having the same mean $\mu_D$ and width $\sigma_D$, and so we minimise the fractional difference. We also want to sample regions in which the velocity dispersion does not vary significantly with radius so that tracers probe similar velocity distributions, which is given by the third term. The set of terms relating to tracers given by the 3$^{\rm rd}$ line in \Eqref{eqn:fitness} maximises the number of orbiting objects, $N_{\rm orb}$, relative to the number of interlopers $N_{\rm int}$ -- that is objects that have not even had a pericentric passage -- and minimise the number of interlopers relative to the number of objects in the window $N_{\rm win}$. The next line maximises the number of tracers in the window and the number of haloes that have more than three objects within the window. Finally we also minimise the halo-to-halo scatter in the LOS velocity dispersion measured using tracers, so as to produce a window that has little scatter in the difference between $\sigma_{\rm LOS}$ \& $\sigma_{v,{\rm H}}$. 

\par
\begin{figure}
    \centering
    \includegraphics[width=0.49\textwidth]{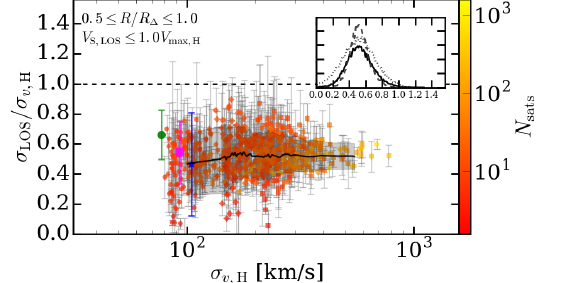}
    \caption{{\bf Observed Satellite/Halo Dispersions with Cuts}. Similar to bottom panel of \Figref{fig:haloobsdisp} but where we have applied cuts to clean the candidate tracer catalogue.}
    \label{fig:haloobsdispcuts}
\end{figure}

 
We find that the optimal window is centred on $D_p\approx0.75R_{\rm vir}$ with a width of $\Delta D_p=0.5R_{\rm vir}$, and $v_{\rm LOS,max}=1.0V_{\rm max}\approx0.5V_{\rm esc}$. This window introduces a bias since it is dominated by objects with small LOS motions, underestimating the dispersion. This bias is seen in \Figref{fig:haloobsdispcuts}, where we have applied these cuts. The bias here is significant, $\sigma_{{\rm LOS}}$ underestimates the true dispersion by $\sim0.5$. We fit the distributions as seen in the inset with Gaussians and find $(\mu,\sigma)=(0.51,0.10)$ for halo dispersion of $\sigma_{v,{\rm H}}\gtrsim250$~km/s. The bias is independent of halo dispersion (mass), although the scatter increases slightly with decreasing halo mass up to $0.2$ for $80~{\rm km/s}\lesssim\sigma_{v,{\rm H}}\lesssim150~$km/s (in part due to decreasing numbers of satellites). The scatter in this ratio arising from variations in lines-of-sight is $\approx0.07$. Applying a similar selection cut to particles from our idealised realisations of haloes gives a bias of $0.55\pm0.2$. This indicates these cuts results in the same dispersion measurement one would expect for true tracers of the halo potential.

\subsection{MW, M31 and LG}
It is worth noting that the dispersions plotted in \Figref{fig:haloobsdisp} (and the projected distance cuts used obtaining these dispersions) are with respect to a well defined barycentre, analogues to measurements made in spectroscopic surveys and not necessarily those made in estimating MW, M31 and LG dispersions. We discuss each of these caveats in turn, although, we argue that this bias also explain the low velocity dispersion measured by \cite{mcconnachie2012a} for the Milky Way, M31 and the Local Group (LG) systems (plotted in \Figref{fig:haloobsdisp} \& \Figref{fig:haloobsdispcuts}) as the effective selection cuts used in \cite{mcconnachie2012a} to measure the dispersion of the Local Group are similar to the one we propose here. 

\par 
First, 3D radial cuts are used, not projected radial cuts. However, so long as subhalos within a small projected distance are removed, the resulting 3D distribution is similar to that which would result from a 3D cut. For LG, dispersion reported in \cite{mcconnachie2012a} uses satellites that are within 3D distance of $\approx2R_{\rm vir}$ (assuming a mass of $\sim5\times10^{12}\Msun$). Our optimal projected distance cut includes objects out to these distance. However, objects at these distances should be treated with caution as those with radial velocities $\gtrsim0.25\times V_{\rm circ}$ tend to be on first infall or not orbiting the host (see \Figref{fig:satelliteorbitsstate}). Overall, these 3D radial cuts do not greatly affect the biased dispersion. 

\par 
While M31 observations are more in keeping with LOS measurements presented in \Figref{fig:haloobsdispcuts}, the MW LOS velocities are nearly-radial velocity wrt to the centre of MW. We find using radial velocities relative to the central halo instead of a uniform LOS introduces no significant bias. When using all objects within 3D distances of $\sim1-2R_{\rm vir}$, we find $\sigma_{\rm LOS}=0.92\pm0.25\sigma_{\rm R}$, where $\sigma_{\rm R}$ is the radial velocity dispersion. Placing a tighter radial cut does not change this relation significantly. Thus, the bias is present in MW observations using radial velocities.

\par 
For the LG system, the velocities are neither LOS velocities nor radial velocities wrt the barycentre. Observations only measure radial velocities of satellites wrt to MW and this must be correct to the LG barycentre frame. This correction is done by removing the velocity of MW relative to LG from velocities of each satellite and using the resulting velocity in the direction of the barycentre to calculate dispersions. Another caveat to consider is the fact that the LG system is not a single virialised halo, but an early stage merger. 

\par
Let's address the issue of mergers first. For late-stage mergers, that is one where the largest subhalo of the host dark matter halo is $>0.5$ times mass of host, we find simply using an arbitrary LOS gives $\sigma_{\rm LOS}=0.58\pm0.17\sigma_{v,{\rm H}}$. For early stage mergers, that is one where two dark matter haloes are infalling but separated by $\gtrsim1.5R_{\rm vir}$ and are of similar masses, $\sigma_{\rm LOS}=0.55\pm0.17\sigma_{v,{\rm H}}$ (assuming a dispersion based on the combined masses of the two merging haloes using \Eqref{eqn:sigmamass}). Thus, the relation holds reasonably well even for early stage mergers. 

\par
\begin{figure}
    \centering
    \includegraphics[width=0.49\textwidth]{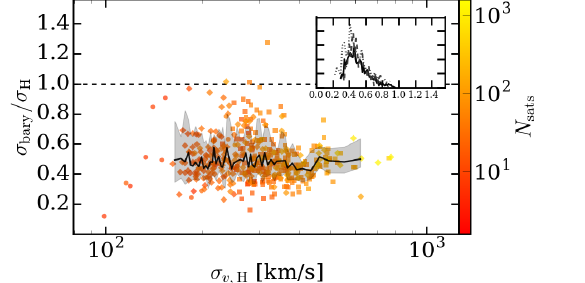}
    \caption{{\bf Observed Satellite About Barycentre of LG Analogues}. Similar to bottom panel of \Figref{fig:haloobsdisp} but where we have applied cuts and mimicked LG observations for early stage merging haloes.}
    \label{fig:LGobs}
\end{figure}

 
If we additionally mimic LG observations, i.e., using LOS velocities wrt to the primary (MW) in the direction of the barycentre (LG) while accounting for the motion of the primary towards the barycentre, the result is also biased. We find a barycentre dispersion to halo dispersion relation of $\sigma_{\rm bary}=0.44\pm0.14\sigma_{v,{\rm H}}$ as seen in \Figref{fig:LGobs}. Although we do not have many merging systems at LG mass scales, the bias show no dependence on mass and the expectation is that the LG dispersion underestimates the merging system's dispersion.

\par 
Though difficult to optimise the selection criteria for merging systems, we argue against using satellites with large positive velocities located $\gtrsim1.5R_{\rm vir}$ away, like Tucana, as this portion of phase-space is dominated by unassociated objects (see orbital state plot in \Figref{fig:satelliteorbitsstate}). If we remove all outward going objects with velocities of $v_{\rm LG}\gtrsim \left(0.1V_{\rm MW,escape}\right)$, which at $2R_{\rm vir}\approx42$~km/s (assuming $V_{\rm MW,circ}$ follows an NFW profile with a density concentration of $c\approx10$, $R_{\rm vir}\approx500$~kpc and a maximum circular velocity of $\approx350$~km/s) and keep objects at distances $\gtrsim600$~kpc\footnote{This list includes Aquarius, SagDIR, UGC4879, LeoA, WLM, LeoT, PegDIR and Cetus.}, we get a mean velocity of $\left\langle V_{\rm LG}\right\rangle=-7\pm17$~km/s and $\sigma_{\rm LG}=48\pm35$~km/s, giving a true halo dispersion of $\sim100$~km/s.

\section{Discussion \& Conclusion}\label{sec:discussion} 
Using the {\small{SURFS}} suite that provides a sample of tens of thousands of (sub)haloes with accurate, well-tracked orbits, we have explored the relationship between a dark matter halo's intrinsic, 3D velocity dispersion, which is a reliable proxy for its mass, and the LOS velocity dispersion deduced from subhaloes (satellite galaxies). In particular, we have analysed the key assumption, namely that neighbouring galaxies are satellite galaxies orbiting the central and are therefore reliable tracers of the underlying halo's phase-space distribution. We showed that the average halo velocity dispersion is tightly correlated with halo mass, with a scatter of $7\%$, but the same is not true when using the LOS motions of surrounding haloes within a three dimensional distance of $\lesssim2R_{\rm vir}$. 

\par 
Comparing the LOS dispersion to the true halo dispersions, we find it underestimates the dispersion on average but has significant scatter. The large scatter is a result of the orbits sampled when uniformly sampling the projected phase-space populated by (sub)haloes surrounding a larger central halo (cf. \Figref{fig:satelliteorbitslos}). Much of the phase-space is dominated by interlopers or newly infalling haloes that are not tracers of the halo's phase-space. 

\par 
Our data allow us to identify the optimal window in projected distance and relative LOS velocity where dispersion measurements should be made to {\em minimise the scatter in the $\sigma_{\rm LOS}-\sigma_{v,{\rm H}}$ relation} and keep any systematic bias mass independent. Applying a projected distance cut of $0.5\lesssim D_p/R_{\rm vir}\lesssim1.0$, where $R_{\rm vir}$ is the virial radius and a LOS velocity cut of $\lesssim0.5V_{\rm esc}$, where $V_{\rm esc}$ is the escape velocity, ensures that one has a significant fraction of orbiting subhaloes, some newly infalling objects and few interlopers. The resulting LOS dispersion $\sigma_{\rm LOS}$ is a mass-independent, biased estimate of the true  $\sigma_{v,{\rm H}}$, with a small amount of scatter: $\sigma_{\rm LOS}=(0.5\pm0.1)\sigma_{v,{\rm H}}$. The small scatter means that LOS dispersion measurements within this window can be trivially corrected to produce halo dispersion and hence virial halo masses. 

\par 
Approximately $60\%$ of all haloes in our catalogue have at least three haloes within this window, with most haloes in our catalogue being low mass groups. The typical satellites had accretion masses of $\log M_{\rm vir}/\Msun=10.9^{+0.6}_{-0.5}$, i.e., catalogues need to be complete to stellar masses of $\sim10^{8}\Msun$. For higher stellar mass cuts ($M_*\gtrsim10^{9}\Msun,M_{\rm vir}\gtrsim10^{11}\Msun$) the fraction of host haloes for which a dispersion can be measured within this optimal window drops significantly. Only $27\%$ of groups with $M_{\rm vir}\gtrsim10^{13}\Msun$ have at least three satellites with masses $\geq10^{11}\Msun$, though by $M_{\rm vir}\gtrsim10^{13.5}\Msun$ this percentage increases to $82\%$. Introducing a larger window naturally increases the haloes for which a $\sigma_{\rm LOS}$ can be measured but these should be flagged as low fidelity estimates. 

\par 
Observational group catalogues, like \cite{yang2006,robotham2011a}, or cluster mass estimates, like those compared in \cite{old2015a,old2017a}, could be improved using this window, producing a high fidelity catalogue with mass uncertainties of $\lesssim0.5$~dex \cite[similar to the uncertainty reported in][who used far more detailed modelling, albeit making numerous assumptions and requiring full 3D velocities]{li2017ApJ...850..116L}. Groups and clusters with more than 10 members would have significantly reduced error bars and methods that use iterative cleaning to remove interlopers would benefit from using such a window. 

\par 
This bias also naturally explains the results from \cite{mcconnachie2012a}, who effectively used similar cuts when estimating dispersions. Mimicking observations gives a bias of $\sigma_{\rm LOS}=(0.44\pm0.14)\sigma_{v,{\rm H}}$. No longer is the Local Group unusually cold but instead lies comfortably within the $1\sigma$ scatter. Using our LG dispersion with the LOS correction or using the LG analogues correction, we predict a halo velocity dispersion of $95\pm72$~km/s or $110\pm88$~km/s respectively. The resulting LG mass is $\log M_{\rm LG, vir}/M_\odot=11.83^{+0.73}_{-1.84}$ or $12.02^{+0.76}_{-2.08}$, far more reasonable than the $\sim1\times10^{11}\Msun$ one would get with the observed LOS dispersion of $49$~km/s (and consistent with $\sim2\times10^{12}\Msun$ to within $1\sigma$, in agreement with historical estimates such as \citealp{courteau1999a}). The corrected dispersion of the MW and M31 systems are similar giving masses of $\log M_{\rm MW,vir}/M_\odot=12.04^{+0.36}_{-0.50}$ and $\log M_{\rm M31,vir}/M_\odot=12.03^{+0.47}_{-0.75}$ respectively, with the MW mass in better agreement with the lower mass estimates of the MW from other studies. 

\section*{Acknowledgements}
The authors would like to acknowledge Geraint Lewis for highlighting the measured LG dispersion and useful discussions. We thank the referee, Alan McConnachie for useful comments. 

\par 
PJE is supported by the Australia Research Council (ARC) Discovery Project Grant DP160102235 and ARC Centre of Excellence ASTRO 3D through project number CE170100013.
CP is supported by ARC Future Fellowship FT130100041. 
CL is funded by a Discovery Early Career Researcher Award DE150100618. 
CL also thanks the MERAC Foundation for a Postdoctoral Research Award. 
CP and AR acknowledge the support of ARC Discovery Project grant DP140100395. 
RP is supported by a University of Western Australia Scholarship. 
Parts of this research were conducted by the ARC Centre of Excellence for All-sky Astrophysics (CAASTRO), through project number CE110001020, and supported by the ARC Discovery Project DP160102235. This research was undertaken on Magnus at the Pawsey Supercomputing Centre in Perth, Australia and on Raijin, the NCI National Facility in Canberra, Australia, which is supported by the Australian commonwealth Government.

\par
The authors contributed to this paper in the following ways: PJE ran simulations and analysed the data, made the plots and wrote the bulk of the paper. CP ran simulations and wrote sections of the paper. All authors have read and commented on the paper.

\vspace{-10pt}
\paragraph*{Facilities} Magnus (Pawsey Supercomputing Centre), Raijin (NCI National Facility)
\vspace{-10pt}
\paragraph*{Software} Python, Matplotlib \cite[][]{matplotlib}, Scipy \cite[][]{scipy}, SciKit \cite[][]{scikit}, Gadget \cite[][]{gadget2}, VELOCIraptor \cite[][Elahi et al, in prep]{elahi2011}, TreeFrog (Elahi et al, in prep), OrbWeaver (Elahi et al, in prep)

\pdfbookmark[1]{References}{sec:ref}
\bibliographystyle{mn2e}
\bibliography{galgroups.bbl}

\begin{thebibliography}{50}
\expandafter\ifx\csname natexlab\endcsname\relax\def\natexlab#1{#1}\fi

\bibitem[{{Barber} {et~al}\mbox{.}(2014){Barber}, {Starkenburg}, {Navarro},
  {McConnachie}, \& {Fattahi}}]{barber2014a}
{Barber} C., {Starkenburg} E., {Navarro} J.~F., {McConnachie} A.~W., {Fattahi}
  A., 2014, \mnras, 437, 959

\bibitem[{{Baugh}(2006)}]{baugh2006}
{Baugh} C.~M., 2006, Reports on Progress in Physics, 69, 3101

\bibitem[{{Behroozi} {et~al}\mbox{.}(2010){Behroozi}, {Conroy}, \&
  {Wechsler}}]{behroozi2010a}
{Behroozi} P.~S., {Conroy} C., {Wechsler} R.~H., 2010, \apj, 717, 379

\bibitem[{{Benson}(2010)}]{benson2010b}
{Benson} A.~J., 2010, \physrep, 495, 33

\bibitem[{{Besla} {et~al}\mbox{.}(2007){Besla}, {Kallivayalil}, {Hernquist},
  {Robertson}, {Cox}, {van der Marel}, \& {Alcock}}]{besla2007a}
{Besla} G., {Kallivayalil} N., {Hernquist} L., {Robertson} B., {Cox} T.~J.,
  {van der Marel} R.~P., {Alcock} C., 2007, \apj, 668, 949

\bibitem[{{Carlberg} {et~al}\mbox{.}(1997){Carlberg}, {Yee}, {Ellingson},
  {Morris}, {Abraham}, {Gravel}, {Pritchet}, {Smecker-Hane}, {Hartwick},
  {Hesser}, {Hutchings}, \& {Oke}}]{carlberg1997b}
{Carlberg} R.~G. {et~al.}, 1997, \apjl, 476, L7

\bibitem[{{Courteau} {et~al}\mbox{.}(2014){Courteau}, {Cappellari}, {de Jong},
  {Dutton}, {Emsellem}, {Hoekstra}, {Koopmans}, {Mamon}, {Maraston}, {Treu}, \&
  {Widrow}}]{courteau2014a}
{Courteau} S. {et~al.}, 2014, Reviews of Modern Physics, 86, 47

\bibitem[{{Courteau} \& {van den Bergh}(1999)}]{courteau1999a}
{Courteau} S., {van den Bergh} S., 1999, \aj, 118, 337

\bibitem[{{Eadie} \& {Harris}(2016)}]{eadie2016a}
{Eadie} G.~M., {Harris} W.~E., 2016, \apj, 829, 108

\bibitem[{{Eadie} {et~al}\mbox{.}(2017){Eadie}, {Springford}, \&
  {Harris}}]{eadie2017a}
{Eadie} G.~M., {Springford} A., {Harris} W.~E., 2017, \apj, 835, 167

\bibitem[{{Elahi} {et~al}\mbox{.}(2013){Elahi}, {Han}, {Lux}, {Ascasibar},
  {Behroozi}, {Knebe}, {Muldrew}, {Onions}, \& {Pearce}}]{elahi2013a}
{Elahi} P.~J. {et~al.}, 2013, \mnras, 433, 1537

\bibitem[{{Elahi} {et~al}\mbox{.}(2011){Elahi}, {Thacker}, \&
  {Widrow}}]{elahi2011}
{Elahi} P.~J., {Thacker} R.~J., {Widrow} L.~M., 2011, \mnras, 418, 320

\bibitem[{{Elahi} {et~al}\mbox{.}(2018){Elahi}, {Welker}, {Power}, {del P
  Lagos}, {Robotham}, {Ca{\~n}as}, \& {Poulton}}]{elahi2018a}
{Elahi} P.~J., {Welker} C., {Power} C., {del P Lagos} C., {Robotham} A.~S.~G.,
  {Ca{\~n}as} R., {Poulton} R., 2018, \mnras

\bibitem[{{Garrison-Kimmel} {et~al}\mbox{.}(2017){Garrison-Kimmel}, {Wetzel},
  {Bullock}, {Hopkins}, {Boylan-Kolchin}, {Faucher-Gigu{\`e}re}, {Kere{\v s}},
  {Quataert}, {Sanderson}, {Graus}, \& {Kelley}}]{garrison-kimmel2017a}
{Garrison-Kimmel} S. {et~al.}, 2017, \mnras, 471, 1709

\bibitem[{{Gill} {et~al}\mbox{.}(2005){Gill}, {Knebe}, \& {Gibson}}]{gill2005}
{Gill} S.~P.~D., {Knebe} A., {Gibson} B.~K., 2005, \mnras, 356, 1327

\bibitem[{{Guo} {et~al}\mbox{.}(2010){Guo}, {White}, {Li}, \&
  {Boylan-Kolchin}}]{guo2010a}
{Guo} Q., {White} S., {Li} C., {Boylan-Kolchin} M., 2010, \mnras, 404, 1111

\bibitem[{{Huang} {et~al}\mbox{.}(2016){Huang}, {Liu}, {Yuan}, {Xiang},
  {Zhang}, {Chen}, {Ren}, {Wang}, {Zhang}, {Hou}, {Wang}, \&
  {Cao}}]{huang2016a}
{Huang} Y. {et~al.}, 2016, \mnras, 463, 2623

\bibitem[{Hunter(2007)}]{matplotlib}
Hunter J.~D., 2007, Computing In Science \& Engineering, 9, 90

\bibitem[{{Jaff{\'e}} {et~al}\mbox{.}(2015){Jaff{\'e}}, {Smith}, {Candlish},
  {Poggianti}, {Sheen}, \& {Verheijen}}]{jaffe2015a}
{Jaff{\'e}} Y.~L., {Smith} R., {Candlish} G.~N., {Poggianti} B.~M., {Sheen}
  Y.-K., {Verheijen} M.~A.~W., 2015, \mnras, 448, 1715

\bibitem[{Jones {et~al}\mbox{.}(2001--)Jones, Oliphant, Peterson,
  {et~al.}}]{scipy}
Jones E., Oliphant T., Peterson P., {et~al.}, 2001--, {SciPy}: Open source
  scientific tools for {Python}. [Online; accessed <today>]

\bibitem[{{Kafle} {et~al}\mbox{.}(2014){Kafle}, {Sharma}, {Lewis}, \&
  {Bland-Hawthorn}}]{kafle2014a}
{Kafle} P.~R., {Sharma} S., {Lewis} G.~F., {Bland-Hawthorn} J., 2014, \apj,
  794, 59

\bibitem[{{Li} {et~al}\mbox{.}(2017){Li}, {Jing}, {Qian}, {Yuan}, \&
  {Zhao}}]{li2017ApJ...850..116L}
{Li} Z.-Z., {Jing} Y.~P., {Qian} Y.-Z., {Yuan} Z., {Zhao} D.-H., 2017, \apj,
  850, 116

\bibitem[{{McConnachie}(2012)}]{mcconnachie2012a}
{McConnachie} A.~W., 2012, \aj, 144, 4

\bibitem[{{McMillan}(2017)}]{mcmillan2017a}
{McMillan} P.~J., 2017, \mnras, 465, 76

\bibitem[{{Moster} {et~al}\mbox{.}(2010){Moster}, {Somerville}, {Maulbetsch},
  {van den Bosch}, {Macci{\`o}}, {Naab}, \& {Oser}}]{moster2010a}
{Moster} B.~P., {Somerville} R.~S., {Maulbetsch} C., {van den Bosch} F.~C.,
  {Macci{\`o}} A.~V., {Naab} T., {Oser} L., 2010, \apj, 710, 903

\bibitem[{{Old} {et~al}\mbox{.}(2014){Old}, {Skibba}, {Pearce}, {Croton},
  {Muldrew}, {Mu{\~n}oz-Cuartas}, {Gifford}, {Gray}, {der Linden}, {Mamon},
  {Merrifield}, {M{\"u}ller}, {Pearson}, {Ponman}, {Saro}, {Sepp}, {Sif{\'o}n},
  {Tempel}, {Tundo}, {Wang}, \& {Wojtak}}]{old2014a}
{Old} L. {et~al.}, 2014, \mnras, 441, 1513

\bibitem[{{Old} {et~al}\mbox{.}(2015){Old}, {Wojtak}, {Mamon}, {Skibba},
  {Pearce}, {Croton}, {Bamford}, {Behroozi}, {de Carvalho},
  {Mu{\~n}oz-Cuartas}, {Gifford}, {Gray}, {der Linden}, {Merrifield},
  {Muldrew}, {M{\"u}ller}, {Pearson}, {Ponman}, {Rozo}, {Rykoff}, {Saro},
  {Sepp}, {Sif{\'o}n}, \& {Tempel}}]{old2015a}
{Old} L. {et~al.}, 2015, \mnras, 449, 1897

\bibitem[{{Old} {et~al}\mbox{.}(2017){Old}, {Wojtak}, {Pearce}, {Gray},
  {Mamon}, {Sif{\'o}n}, {Tempel}, {Biviano}, {Yee}, {de Carvalho},
  {M{\"u}ller}, {Sepp}, {Skibba}, {Croton}, {Power}, {von der Linden}, \&
  {Saro}}]{old2017a}
{Old} L. {et~al.}, 2017, ArXiv e-prints

\bibitem[{{Oman} \& {Hudson}(2016)}]{oman2016b}
{Oman} K.~A., {Hudson} M.~J., 2016, \mnras, 463, 3083

\bibitem[{{Oman} {et~al}\mbox{.}(2013){Oman}, {Hudson}, \&
  {Behroozi}}]{oman2013a}
{Oman} K.~A., {Hudson} M.~J., {Behroozi} P.~S., 2013, \mnras, 431, 2307

\bibitem[{Pedregosa {et~al}\mbox{.}(2011)Pedregosa, Varoquaux, Gramfort,
  Michel, Thirion, Grisel, Blondel, Prettenhofer, Weiss, Dubourg, Vanderplas,
  Passos, Cournapeau, Brucher, Perrot, \& Duchesnay}]{scikit}
Pedregosa F. {et~al.}, 2011, Journal of Machine Learning Research, 12, 2825

\bibitem[{{Phelps} {et~al}\mbox{.}(2013){Phelps}, {Nusser}, \&
  {Desjacques}}]{phelps2013a}
{Phelps} S., {Nusser} A., {Desjacques} V., 2013, \apj, 775, 102

\bibitem[{{Planck Collaboration} {et~al}\mbox{.}(2015){Planck Collaboration},
  {Ade}, {Aghanim}, {Arnaud}, {Ashdown}, {Aumont}, {Baccigalupi}, {Banday},
  {Barreiro}, {Bartlett}, \& et~al.}]{planckcosmoparams2015}
{Planck Collaboration} {et~al.}, 2015, ArXiv e-prints

\bibitem[{{Reddick} {et~al}\mbox{.}(2013){Reddick}, {Wechsler}, {Tinker}, \&
  {Behroozi}}]{reddick2013a}
{Reddick} R.~M., {Wechsler} R.~H., {Tinker} J.~L., {Behroozi} P.~S., 2013,
  \apj, 771, 30

\bibitem[{{Reid} \& {Dame}(2016)}]{reid2016a}
{Reid} M.~J., {Dame} T.~M., 2016, \apj, 832, 159

\bibitem[{{Robotham} {et~al}\mbox{.}(2011){Robotham}, {Norberg}, {Driver},
  {Baldry}, {Bamford}, {Hopkins}, {Liske}, {Loveday}, {Merson}, {Peacock},
  {Brough}, {Cameron}, {Conselice}, {Croom}, {Frenk}, {Gunawardhana}, {Hill},
  {Jones}, {Kelvin}, {Kuijken}, {Nichol}, {Parkinson}, {Pimbblet}, {Phillipps},
  {Popescu}, {Prescott}, {Sharp}, {Sutherland}, {Taylor}, {Thomas}, {Tuffs},
  {van Kampen}, \& {Wijesinghe}}]{robotham2011a}
{Robotham} A.~S.~G. {et~al.}, 2011, \mnras, 416, 2640

\bibitem[{{Rodr{\'{\i}}guez-Puebla}
  {et~al}\mbox{.}(2016){Rodr{\'{\i}}guez-Puebla}, {Behroozi}, {Primack},
  {Klypin}, {Lee}, \& {Hellinger}}]{rodriguezpuebla2016a}
{Rodr{\'{\i}}guez-Puebla} A., {Behroozi} P., {Primack} J., {Klypin} A., {Lee}
  C., {Hellinger} D., 2016, \mnras, 462, 893

\bibitem[{{Sawala} {et~al}\mbox{.}(2015){Sawala}, {Frenk}, {Fattahi},
  {Navarro}, {Bower}, {Crain}, {Dalla Vecchia}, {Furlong}, {Jenkins},
  {McCarthy}, {Qu}, {Schaller}, {Schaye}, \& {Theuns}}]{sawala2015a}
{Sawala} T. {et~al.}, 2015, \mnras, 448, 2941

\bibitem[{{Sawala} {et~al}\mbox{.}(2017){Sawala}, {Pihajoki}, {Johansson},
  {Frenk}, {Navarro}, {Oman}, \& {White}}]{sawala2017a}
{Sawala} T., {Pihajoki} P., {Johansson} P.~H., {Frenk} C.~S., {Navarro} J.~F.,
  {Oman} K.~A., {White} S.~D.~M., 2017, \mnras, 467, 4383

\bibitem[{{Schneider}(2006)}]{schneider2006}
{Schneider} P., 2006, {Extragalactic Astronomy and Cosmology}

\bibitem[{{Springel}(2005)}]{gadget2}
{Springel} V., 2005, \mnras, 364, 1105

\bibitem[{{van den Bosch}(2017)}]{vandenbosch2017a}
{van den Bosch} F.~C., 2017, \mnras, 468, 885

\bibitem[{{van Uitert} {et~al}\mbox{.}(2016){van Uitert}, {Cacciato},
  {Hoekstra}, {Brouwer}, {Sif{\'o}n}, {Viola}, {Baldry}, {Bland-Hawthorn},
  {Brough}, {Brown}, {Choi}, {Driver}, {Erben}, {Heymans}, {Hildebrandt},
  {Joachimi}, {Kuijken}, {Liske}, {Loveday}, {McFarland}, {Miller}, {Nakajima},
  {Peacock}, {Radovich}, {Robotham}, {Schneider}, {Sikkema}, {Taylor}, \&
  {Verdoes Kleijn}}]{vanuitert2016a}
{van Uitert} E. {et~al.}, 2016, \mnras, 459, 3251

\bibitem[{{Watkins} {et~al}\mbox{.}(2010){Watkins}, {Evans}, \&
  {An}}]{watkins2010}
{Watkins} L.~L., {Evans} N.~W., {An} J.~H., 2010, \mnras, 406, 264

\bibitem[{{White} \& {Frenk}(1991)}]{white1991a}
{White} S.~D.~M., {Frenk} C.~S., 1991, \apj, 379, 52

\bibitem[{{Widrow} \& {Dubinski}(2005)}]{galactics}
{Widrow} L.~M., {Dubinski} J., 2005, \apj, 631, 838

\bibitem[{{Widrow} {et~al}\mbox{.}(2008){Widrow}, {Pym}, \&
  {Dubinski}}]{widrow2008}
{Widrow} L.~M., {Pym} B., {Dubinski} J., 2008, \apj, 679, 1239

\bibitem[{{Yang} {et~al}\mbox{.}(2006){Yang}, {van den Bosch}, {Mo}, {Mao},
  {Kang}, {Weinmann}, {Guo}, \& {Jing}}]{yang2006}
{Yang} X., {van den Bosch} F.~C., {Mo} H.~J., {Mao} S., {Kang} X., {Weinmann}
  S.~M., {Guo} Y., {Jing} Y.~P., 2006, \mnras, 369, 1293

\bibitem[{{Yoon} {et~al}\mbox{.}(2017){Yoon}, {Chung}, {Smith}, \&
  {Jaff{\'e}}}]{yoon2017a}
{Yoon} H., {Chung} A., {Smith} R., {Jaff{\'e}} Y.~L., 2017, \apj, 838, 81

\bibitem[{{Zaritsky} \& {Courtois}(2017)}]{zaritsky2017a}
{Zaritsky} D., {Courtois} H., 2017, \mnras, 465, 3724

\end{thebibliography}
\appendix
\section{OrbWeaver}\label{sec:orbweaver}
{\sc OrbWeaver} is part of the {\sc VELOCIraptor} tool-kit. It traces the relative motions of (sub)haloes around other haloes using the halo merger tree of {\sc TreeFrog} combined with the halo catalogues of {\sc VELOCIraptor}. This python code (soon to be translated to c++ and make use of the MPI API) calculates a variety of orbital properties, from positions of apsides, ellipticity, orbital period, orbital angular momentum, etc. Orbiting haloes are traced forwards and backwards in time along the merger tree to identify changes in the sign of the radial velocity corresponding to pericentric and apocentric passages. First pericentric passage is defined as the first from negative radial velocities to positive radial velocities that occurs within $2R_{\rm vir}$. We determine the apsides by linearly interpolating the orbiting halo's relative position and velocities between the transition points. Due to the high cadence of our halo catalogue, linear interpolation is a reasonable approximation, though quadratic interpolation of positions is possible. 

\section{Orbits in 3D}\label{sec:orb3d}
Orbital state information as a function of 3D distance from halo centre.
\begin{figure}
    \centering
    \includegraphics[width=0.49\textwidth]{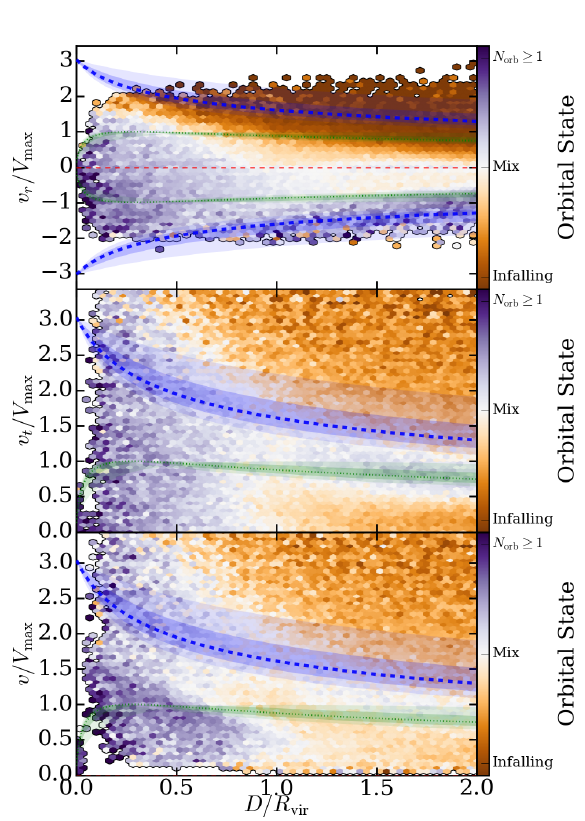}
    \caption{{\bf Orbital state}. We show the typical orbital state (similar to \Figref{fig:satelliteorbitslos}) at a given radial, tangential and total speed distribution as a function of distances (similar to \Figref{fig:satelliteorbits})}
    \label{fig:satelliteorbitsstate}
\end{figure}

\end{document}